\documentclass{article}
\usepackage{amsthm}
\usepackage{amsmath,amssymb} 
\usepackage{newpxtext, newpxmath} 

\makeatletter
\newcommand{\dotminus}{\mathbin{\text{\@dotminus}}}

\newcommand{\@dotminus}{%
  \ooalign{\hidewidth\raise1ex\hbox{.}\hidewidth\cr$\m@th-$\cr}%
}
\makeatother

\usepackage{graphicx}

\title{Quantifying Gerrymandering With Simulated Annealing}

\author{Stuart Wayland \thanks{Department of Computer Science, University of California, Santa Barbara. Research for the completion of the Distinguish in the Major Program. email: stuartwayland@ucsb.edu. Faculty Advisor: Professor Eric Vigoda
}}
\date{August 2022}

\begin{document}
\maketitle
\begin{abstract}
    Gerrymandering is the perversion of an election based on manipulation of voting district boundaries, and has been a historically important yet difficult task to analytically prove.     
    We propose a Markov Chain Monte Carlo with Simulated Annealing as a solution for measuring the extent to which a districting plan is unfair. 
     We put forth promising results in the successful application of redistricting chains for the state of Texas, using an implementation of a redistricting Markov Chain with Simulated Annealing to produce accelerated results. This demonstrates strong evidence that Simulated Annealing is effective in quickly generating representative voting distributions for large elections, and furthermore capable of indicating unfair bias in enacted districting plans. 
\end{abstract}

\section{Introduction}
\par
Legislative redistricting is a necessary and hugely problematic component of democratic processes. It not only plays an integral role in valuing citizens' votes, but also has strong implications in the government's representation of it's people. Unfortunately, the choice of a redistricting, and thus the way we value certain citizens' votes, may be exploited for partisan gain. This notion of manipulating district boundaries to favor an outcome, known as gerrymandering, has become notoriously intertwined with democratic electoral processes. 

\par In order to comment on the existence of gerrymandering in districting plans, we must first understand the procedure in which districtings are assigned. For US Congressional districts, the US Constitution mandates that redistricting occur every 10 years following the census count and reapportionment of seats in the US House of Representatives. Furthermore, the Federal Government requires that each state's congressional districting plans have equal population, are contiguous and compact, and comply with the Voting Rights Act \cite{vra_voting_1965}.  
\par While these requirements are set forth by the Federal Government, each state has an elected board of partisan legislators that determines the enacted plan within these requirements. Should a plan be suspected of being gerrymandered, it is presented to the respective state's supreme court who pass judgement on the fairness of the plan. The reality of judging a plan's fairness, however, has been a highly contested and historically difficult task. In past trials, the efficiency gap has served as the primary indicator for determining a plan's fairness. However, as detailed in Chambers et al. \cite{chambers_flaws_2017}, this metric can yield inaccuracies. We will employ a more robust method of evaluating districting fairness that uses the region's geo-politics to quantify the extent to which a districting is gerrymandered. We propose a Markov Chain Monte Carlo (MCMC) solution to determine districting fairness, as pioneered by Fifield et al. \cite{fifield_automated_2020}, Mattingly's Quantifying Gerrymandering Group at Duke University \cite{herschlag_quantifying_2020}, the MGGG at Tuft's University \cite{deford_redistricting_2019}, and CMU's Wesley Pedgen \cite{pegden_partisan_2017}.
\par 
An MCMC solution provides a well-principled method for handling the vast number of potential districtings that adhere to the Federal governments restrictions on compactness and population equality– we will call these candidates \textit{viable plans}. Barring access to corporation sized computational facilities, enumerating the set of viable plans is computationally ineffective. Instead, we use an MCMC to sample from the space of \textit{viable plans} via a random walk that prefers compact and equal population districtings. We then apply voting data to the collection of generated plans to produce a distribution of feasible election results. 



This distribution provides an associated probability to the enacted plan that represents the likelihood the plan should be chosen. In the case that the enacted plan is gerrymandered, we would expect this plan to have a significantly low probability of being selected. For example, Professor Wesley Pegden showed that the 2010 congressional districting of Pennsylvania gave more seats to Republicans than  99.99\% of his randomly generated viable districting plans, meaning that this plan had a roughly 0.0001 probability of being fairly selected \cite{pegden_partisan_2017}.
\par 
Since the emergence of this field in 2018, the previously mentioned MCMC solution has been presented in court to analyze the presence of gerrymandering. Two of the aforementioned groups provided expert witness testimony. Professor Wesley Pegden of CMU testified in Pennsylvania in 2018, successfully ruling the 2010 Pennsylvania Congressional districting plan as gerrymandered \cite{duffy_pa_2018}. Similarly, Professor Johnathan Mattingly of Duke University provided testimony in North Carolina in 2019, advocating that the 2012 and 2016 North Carolina congressional districting plans were gerrymandered \cite{mattingly_expert_2021}. 

\par 

 We hope to replicate and enhance the experiments and analytical process these groups employed, expanding this application to the state of Texas. We put forth promising results in the successful application of redistricting chains on larger geographic areas while demonstrating a necessity for speed improvements. We proceed to use a modification of Dr. Benjamin Fifield's implementation \cite{kenny_redist_2022} of a redistricting Markov Chain to produce accelerated results using Simulated Annealing, showing strong evidence that this practical method is effective in quickly generating representative voting distributions. 

\section{Background} 
    This section will give a brief overview of the US Census Bureau's geographic division of the United States and explain how the redistricting process can be framed as a graph cut problem, providing context for an MCMC implementation. We hope to: 1) provide an understanding of how random districting samples are generated, and 2) illustrate opportunities to improve districting via a process called Simulated Annealing.

    \par In the scope of an election process, the US Census Bureau divides each state into counties, census tracts, and voting tabulation districts (VTDs). VTDs are the finest geographic division used for aggregating votes, and will be used as the geographic units in the implementation of our MCMC.   
    
    \subsection{Redistricting as a Graph Cut Problem}
        Given a shapefile $S$ that divides a state into geographic units, we consider the dual of the shapefile $G = (V, E)$. The dual $G$ is computed by associating a vertex with each open face in $S$, and an edge between two vertices if their corresponding open faces share a border. Also, we have a function $pop(v): V \to \mathbb{Z}$, where for each $v \in V$, \text{ } $pop(v)$ returns the population of the geographic unit that $v$ represents. 
        
        \par For example, let us consider for simplicity the county shapefile of the state of Iowa and its corresponding dual graph, as shown in \textbf{Figure 1}. In the right image, Iowa's 99 counties are represented by vertices and their borders represented with edges. Moreover, we can extract a county's population using $pop(\cdot)$ on the corresponding vertex.
        
    \begin{figure}[!htbp]
\begin{center} 
\includegraphics[scale = 0.2]{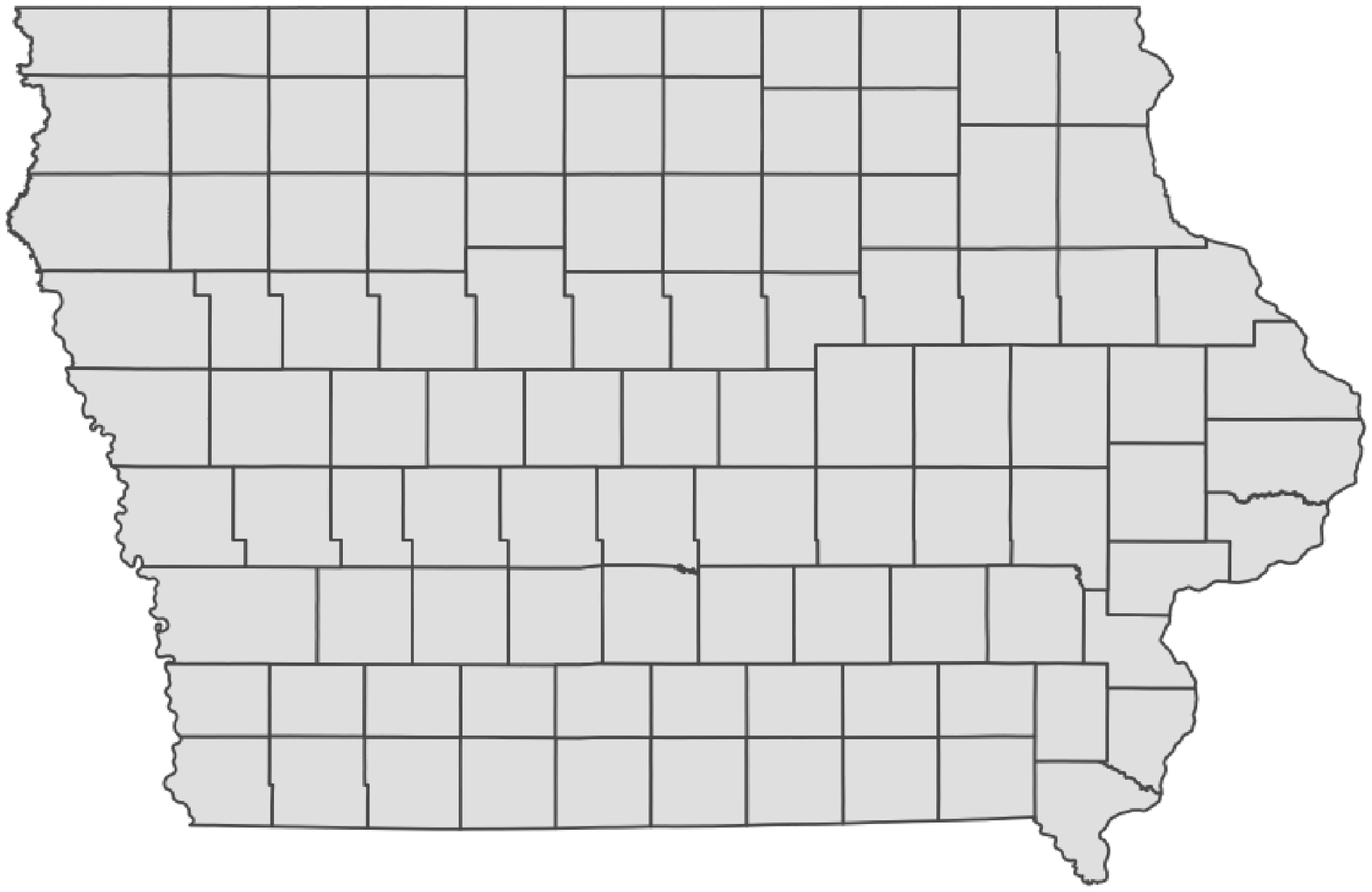}
\includegraphics[scale = 0.2]{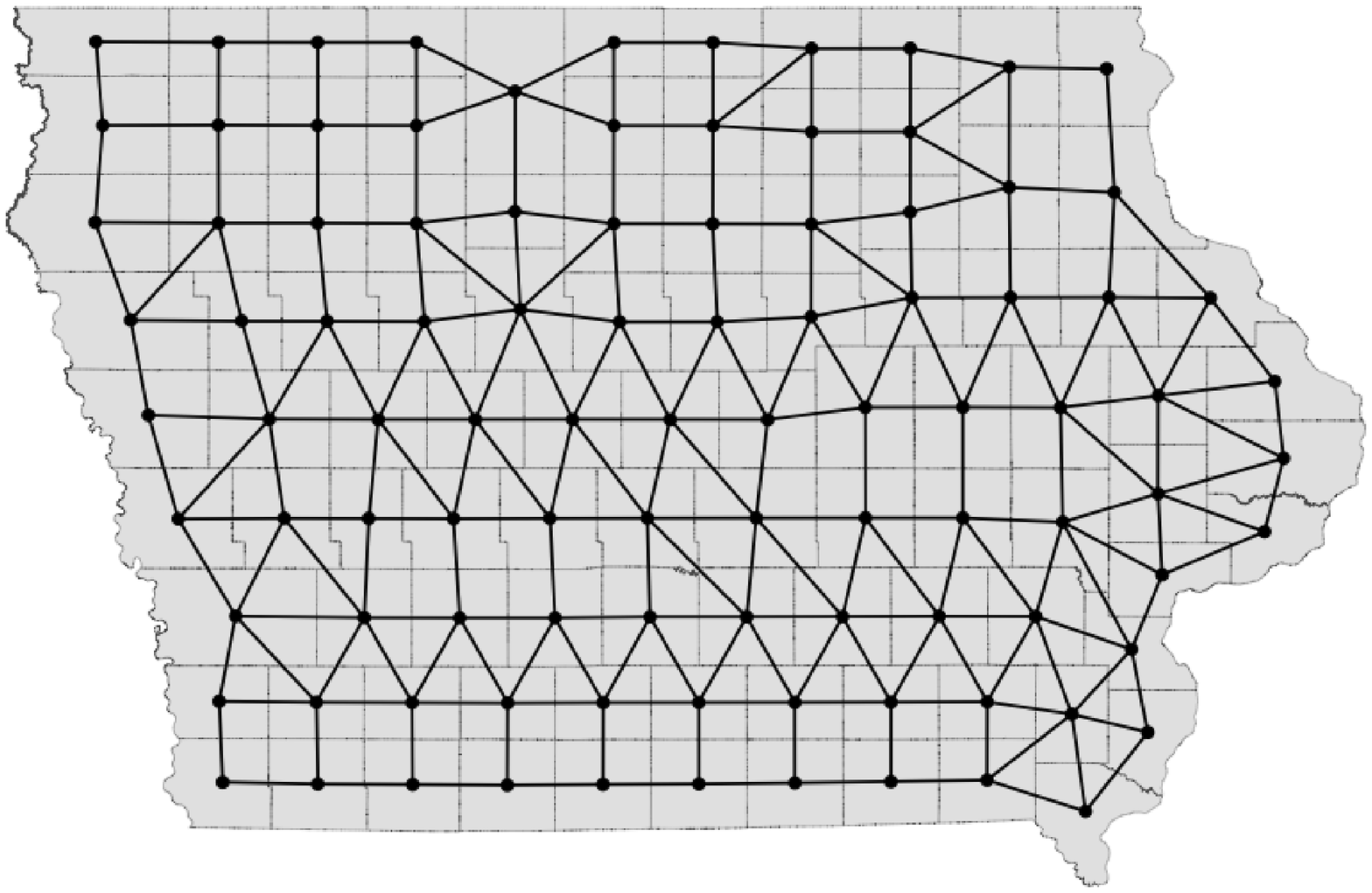}
\end{center}  
\caption{Iowa County Shape and Corresponding Dual}
\end{figure}
\par
In the resulting planar graph, we can represent $n$ congressional districts with an n-coloring. Returning to our Iowa example, we consider representing Iowa's 4 congressional districts by coloring each node of the dual graph with one of four colors, as seen in \textbf{Figure 2}.

\begin{figure}[!htbp]
    \centering
    \includegraphics[scale = 0.2]{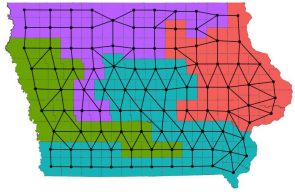}
    \caption{Iowa County Dual with Districting Plan }
    \label{fig:my_label}
\end{figure}
\par
Formally, a \textit{districting plan} $\sigma:V\to\mathbb{N}$ is an n-coloring of a graph $G=(V,E)$. For a districting plan $\sigma$, we may also consider removing all non-Monochromatic edges (edges who have endpoints with different colors) from G, producing a sub-graph that isolates the state's districts. This associates a unique \textit{cut edge set} $S_\sigma$, to each districting $\sigma$, where $S_\sigma\subseteq E$ are the cuts required to separate colored components in the coloring $\sigma$.

\par Federal regulations require that districting plans, and therefore graph colorings, be contiguous. Thus, we can define a \textit{valid plan} as follows: a districting plan $\sigma$ is \textit{valid} if there exists a \textit{cut edge set} $S_\sigma \subseteq E$ such that the sub graph, $(V, E\setminus S_\sigma)$ has exactly $n$ connected components $C_\sigma^1$, ..., $C_\sigma^n$. Let $\Omega$ be the set of all \textit{valid plans}. 
\par These connected components $C_\sigma^1, ..., C_\sigma^n$ are the districts laid out by the plan $\sigma$, and will be referred to as such. Thus, we can calculate the population of a district by aggregating the population of each vertex in the connected component. That is, given a district $C_\sigma^i \subseteq V $, we define the population of $C_\sigma^i$ as $pop(C_\sigma^i) = \sum_{v \in C_\sigma^i} pop(v).$ It then follows that $pop(V)$ is equal to the total population of the state. 
\par Now, we can define two important metrics on a districting plan $\sigma$ and it's \textit{cut edge set} $S_\sigma$. First, the \textit{population equality metric}:
\[
pop\_eq(\sigma) = \sum_{i=1}^n \left | pop(C_\sigma^i) - \frac{pop(V)}{n}\right |
\]

And next, the \textit{compactness metric}, which is determined by the ratio of cut edges to total edges in the dual graph G=(V,E):
\[
comp(\sigma) = \frac{|S_\sigma|}{|E|}.
\]
\par

We hope to favor a smaller subset of $\Omega$ that obey Federal regulations for compactness and population difference. These will be referred to as \textit{viable plans}, and will be targeted by using the score functions $comp()$ and $pop\_eq$.
\par 
We will generate a \textit{viable plan} by performing a random walk on $\Omega$ as follows:
suppose we have a \textit{valid} districting plan $\sigma \in \Omega$, defined by the edge set $S_\sigma$ for the dual graph $G = (V,E)$. We begin by selecting a random vertex $v \in V$ such that $N(v) \cap S_\sigma \not = \emptyset$. We then select a random edge $e \in N(v) \cap S_\sigma$. That is, $v$ is a vertex lying on an internal boundary of a district. We then recolor $v$ to the neighboring district which $v$ is connected to through $e \in N(v) \cap S_\sigma$. Finally, we check that the resulting districting plan from this transition, call it $\sigma^\prime$, belongs to the space of valid districtings, $\Omega$. If $\sigma^\prime$ is not a valid districting, we repeat the process on $\sigma$.

\par 
Once we have produced a new valid districting $\sigma^\prime$, we apply a Metropolis Filter. This will leverage the metrics discussed previously to favor plans with more desirable compactness and population equality properties. This filter sets the probability of transitioning to a new plan equal to the weight of the new plan over the weight of the old. More precisely, for a weight function $w: \Omega \to \mathbb{R}$, we set the transition probability,
\[
p = \min \left \{ \frac{w(\sigma^\prime)}{w(\sigma)}, 1\right \}.
\]
The Metropolis Filter guarantees that our sampling probability distribution is proportional to our weight function $w$, allowing us to control the districting plans we wish to sample. With this in mind, we set our weight function as
\[
w(\sigma) = e^{-(\beta_1 pop\_eq(\sigma) + \beta_2 comp(\sigma))},
\]
where $\beta_1, \beta_2$ are constants. 
We transition to the plan $\sigma^\prime$ with our above probability $p$, setting $\sigma = \sigma^\prime$, or otherwise leave $\sigma$ unchanged. We then repeat the process until we have arrived at a randomly generated districting plan. The number of steps required to achieve a random sample is known as the chain's mixing time \cite{jerrum_markov_1996}, and is a theoretical topic altogether unknown in the scope of redistricting chains. Our goal is to sample from a distribution proportional to $w$ with very large $\beta_1, \beta_2$ values. However, these large $\beta_1, \beta_2$ values severely constrict the mixing of a chain over the space $\Omega$, whereas small $\beta_1, \beta_2$ allow greater movement over space and therefore faster apparent mixing. For this reason, MCMC chains often employ methods to artificially speed up mixing in practice, even if these methods do not ensure any theoretical difference in the mixing time. 

\subsection{Simulated Annealing}
\par 
One such practical method is known as simulated annealing, which dynamically adjusts the $\beta_1,\beta_2$ values as the chain is mixing. 
The $\beta_1, \beta_2$ values control the influence of population equality and the compactness metric respectively. This allows us to manage how severely we wish to enforce our constraint metrics. For example, consider $\beta_1, \beta_2 = 0$, meaning for all $\sigma \in \Omega$,  $w(\sigma) = 1$. This removes all influence of our weight function, resulting in uniformly sampling from our space $\Omega$ of \textit{valid} plans. Conversely, if we consider extremely large $\beta_1, \beta_2$, then our weight function approaches zero for all plans without a zero population equality (all districts have exactly the same population) and compactness score. This results in only sampling extremely compact, equal population districting plans. 
In the context of a normal Markov Chain, we choose our $\beta_1, \beta_2$ values so that our sampled plans fall within federal laws of compactness and population equality. We will refer to the $\beta_1, \beta_2$ values that produce such plans as the target values. 

\par To provide context to the name simulated annealing, we may think of $\beta_1, \beta_2$ values as inverse temperature. Recall that annealing is the process of heating a metal before allowing it to cool slowly in a desired shape. Similarly, simulated annealing refers to beginning a Markov chain with $\beta_1, \beta_2 = 0$. This may be thought of as heating the chain, and is often referred to as the "hot steps", in which the chain may move freely across the entire space of valid districtings. In order to slowly "cool" the  chain, the $\beta_1, \beta_2$ values are increased at a rate called the cooling schedule until they reach their target values. The final steps of the chain at the target $\beta_1, \beta_2$ values are known as the "cold steps", and ensure we are sampling from the desired distribution guaranteed from the Metropolis Filter. 

\par 
The benefit of simulated annealing is its ability to increase the speed of plan generation. This improvement comes from decreasing the total number of steps necessary to produce a near-random plan. Since the majority of the chain is run with small $\beta_1, \beta_2$ values, the probability of accepting a proposed plan in the chain is significantly higher than its pure Markov chain counterpart. This means we may achieve the same number of accepted steps, and so in practice the same amount of "randomness", with far fewer iterations of the chain. 

\par 

\section{Results}
\par We generated randomly sampled districting ensembles for the states of North Carolina and Texas, using North Carolina as a small scale validation study for our simulated annealing algorithm. We compared our results in North Carolina to the findings of Herschlag et al. \cite{herschlag_quantifying_2020}.  
\par We analyze the sets of generated plans, which we will refer to as ensembles, by processing election outcomes for each plan in the ensemble. These election outcomes produce a distribution that we compare to an enacted plan. From the selection of our weight function, we expect to see a near-normal distribution of election outcomes. As such, we fit a density curve to our election outcome data to provide an associated probability to our enacted plan, which we assume to be the probability this outcome should occur fairly. 

\subsection{North Carolina}

\par We begin by sampling an ensemble of congressional districting plans for the 2016 North Carolina US House of Representatives election. We will be using the 2010 North Carolina VTD shapefile, and will be comparing to the North Carolina congressional districting enacted in 2010. 

\par We first replicated the study done by Herschlag et al. \cite{herschlag_quantifying_2020}, producing similar results before testing our simulated annealing algorithm. We generated 50 plans with the Redist flip algorithm to validate our preprocessing and analytic procedures. We compared our distribution of 50 plans to the 66,544 plans generated by Herschlag et al. \cite{herschlag_quantifying_2020} We then fitted density curves to both distributions, as seen in \textbf{Figure 3}.  

\begin{figure}[!htbp]
\begin{center} 
\includegraphics[scale = 0.4]{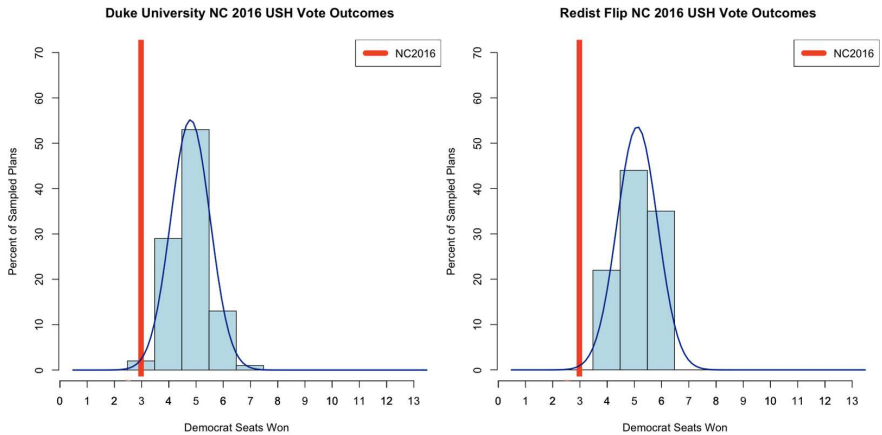}
\end{center}  
\caption{Distribution of number of Democratic seats won among the 13 seats in North Carolina. We display the results of Duke University (left) and our Redist Flip algorithm (right). In both figures, the red line represents the outcome of the enacted plan.}
\end{figure}
\par
Despite only generating 50 plans, we see a similarity to the results in Herschlag et al. \cite{herschlag_quantifying_2020}, who showed plans giving 3 Democratic seats in this election had a probability of 0.008 of being fairly selected. Unsurprisingly, we did not produce a single plan yielding 3 Democratic seats. Regardless of our inadequate number of generated plans, we did validate that our preprocessing of voting/geographic data and our election analysis both align with previous work. We fitted a normal density curve to our plan distribution, calculating a z-score for 3 Democratic seats. We found that a plan yielding a 3 Democratic seat outcome had an approximate probability of 0.0021 of being selected. 

\begin{figure}[!htbp]
\begin{center} 
\includegraphics[scale = 0.4]{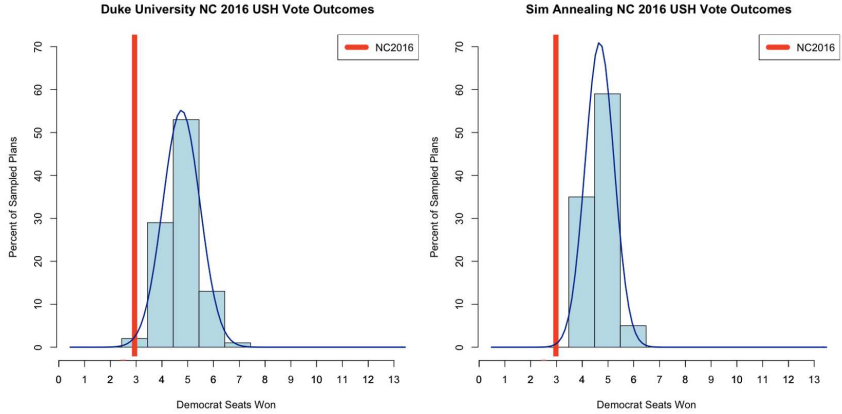}
\end{center}  
\caption{Distribution of number of Democratic seats won among the 13 seats in North Carolina. We display the results of Duke University (left) and our simulated annealing algorithm (right). In both figures, the red line represents the outcome of the enacted plan.}
\end{figure}

After validating our data processing and analysis methods, we generated 500 plans for North Carolina using simulated annealing. As shown in \textbf{Figure 4}, we obtained a similar distribution as Duke, with the vast majority of sampled plans yielding 4, 5, and 6 Democratic seats in Congress, compared to the 3 seats awarded in the 2016 election. Like our flip algorithm, not a single plan generated with simulated annealing produces an outcome of 3 Democratic seats won. Once again, we fitted a normal density curve to our generated distribution to calculate the probabilities of producing an election outcome with 3 Democratic seats won. Our calculated probabilities compared to Mattingly's Duke group\cite{herschlag_quantifying_2020} can be seen in $\textbf{Table 1}$. These results show promise in expanding simulated annealing to a far larger application, namely plan generation for the state of Texas. 
 
 \begin{table}[h!]
\centering
\begin{tabular}{ |c|c|c|c|c|c| } 
 \hline
 Democratic Seats Won &  3 & 4 & 5 & 6 \\
 \hline
 Redistricting Chain & 0.0021 & 0.21 & 0.46 & 0.38 \\
 \hline
 Simulated Annealing & 0.0013 & .33 & .61  & .06\\ 
 \hline
 Duke Group & 0.008 & .29 & .52 & .11 \\
 \hline
\end{tabular}
\caption{Approximate probabilities of 2016 NC USH Election Outcomes from Redistricting chain, Simulated Annealing Distribution and Duke's Quantifying Gerrymandering Group\cite{herschlag_quantifying_2020}}
\end{table}

\subsection{Texas}
\par 
Like in North Carolina, we produce a voting distribution of an ensemble of generated districting plans to analyze outcomes of the 2020 US House of Representatives vote. We will be comparing our generated plans to the 2020 Texas Congressional districting, and again will be producing a distribution from the Redist flip algorithm as well as a distribution generated from simulated annealing. We generated 50 plans using the Redist algorithm and 500 plans using simulated annealing. These two distributions can be seen in \textbf{Figure 5}. 

\begin{figure}[!htbp]
\begin{center} 
\includegraphics[scale = 0.4]{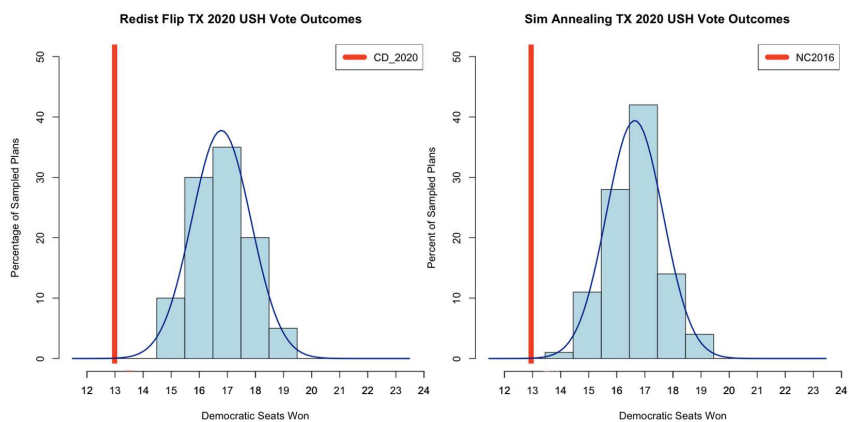}
\end{center}  
\caption{Distribution of number of Democratic seats won in 2020 US House of Representatives election in Texas. Shown left is distribution of 50 plans generated from Redist Flip algorithm, and the right shows distribution of 500 plans generated with simulated annealing. The vertical red line represents the real outcome from the enacted Congressional Districting plan.}
\end{figure}

\par Despite the 2020 Texas Congressional Districting plan yielding 13 of 36 seats to Democrats, our Redist flip algorithm, shown on the left of \textbf{Figure 5}, did not produce a single plan with 13 or even 14 Democratic seats won. The majority of plans generated gave between 16 and 18 seats to Democrats.  \par 
Similarly, our simulated annealing algorithm also did not produce a single plan with the same 13 Democratic seat outcome. Like the flip algorithm, the majority of the 500 generated plans gave between 16 and 18 seats to Democrats. We fitted normal density curves to both distributions, calculating the probability of each election outcome from our simulated annealing distribution, as shown in \textbf{Table 2}.   

\begin{table}[h!]
\centering
\begin{tabular}{ |c|c|c|c|c|c|c|c| } 
 \hline
 Democratic Seats Won &  13 & 14 & 15 & 16 & 17 & 18 & 19 \\ 
 \hline
 Probability & 0.0002 & 0.01 & .11  & .29 & 0.44 & .13 & .03\\ 
 \hline
\end{tabular}
\caption{Approximate probabilities of 2020 TX USH Election Outcomes from Simulated Annealing Distribution }
\end{table}

\section{Methodology}
    \par
    In this section, we describe the proposed methodology for running the Redist package's flip algorithm in the states of North Carolina, Iowa, and Texas. We will provide the specifications of the Markov Chain algorithm used to generate the voting distributions, the preprocessing methods of external shape files and voting data sets, and the precise simulation inputs and reasoning behind the parameter choices. Considering the state-specific data sources, these preprocessing and parameter choices will vary based on the size of the state and the manner in which voting and population data is recorded. Furthermore, we will detail the simulated annealing algorithm used to improve the speed of plan generation, reporting the specifications yielding the results provided above.  
    
\subsection{The Redist Flip Algorithm}
    \par
    The Redist flip algorithm is an expansion of the Swendsen-Wang Algorithm \cite{barbu_generalizing_2005}, a simple Markov Chain described in Section 2. Now, let us set about describing the chain used to generate the above results.  
    \par
    We will then set forth describing the transition on step $t$ of the flip chain, wherein we obtain $\sigma_{t+1}$, defined by edge set $S_{\sigma_{t+1}}$ from $\sigma_t$, defined by edge set $S_{\sigma_t}$.
    \par Seeing how $S_\sigma$ refers to the cut, or non-monochromatic edges in the plan $\sigma$, let us define $\overline{E}(\sigma)$ to be the set of uncut, or monochromatic edges in the plan $\sigma$. That is, 
    
\[
\overline{E}(\sigma):= \left \{ e \in E\backslash S_\sigma \right \}.
\]
The algorithm begins by labelling each edge $e \in \overline{E}(\sigma_t)$ as a "flip" edge with probability $\lambda$. Let us refer to this labelling of edges $e \in \overline{E}(\sigma_t)$ as $\pi$. We then define 
\[
    \overline{E}_{flip}(\sigma_t, \pi) = \left \{ e \in \overline{E}(\sigma_t): \pi(e) = 1 \right \}.
\]
In other words, $ \overline{E}_{flip}(\sigma_t, \pi)$ is the set of "flip" edges resulting from $\pi$. 
 $\overline{E}_{flip}(\sigma_t, \pi) \subseteq \overline{E}(\sigma_t)$ can also be thought of as creating a set of connected components lying within $\overline{E}(\sigma_t)$. Let us call this set of connected components resulting from our labeling of flip edges $C_{flip}(\sigma_t)$. Next, let us consider the set $C_\partial(\sigma_t) \subseteq C_{flip}(\sigma_t)$, which we define as 
    \[
    C_\partial(\sigma_t) := \left \{ C \in C_{flip}(\sigma_t) : \exists v \in C \text{ s.t. } \exists \text{ } e \in N(v) \text{ where } e \in S_{\sigma_t}\right \}.
    \]
    \par
    That is, $C_\partial(\sigma_t)$ is the set of connected components from the "flip" labeling that lie on the boundary of the plan $\sigma_t$. Next, the algorithm selects a random subset $F \subseteq C_\partial(\pi_t)$ such that $F$ does not contain any adjacent connected components, and flips each component within $F$ its respective neighboring coloring. In the case that a component has multiple neighboring districts, a neighboring district is chosen at random. Let us call the resulting plan from this flip of components $\sigma^\prime$.
    \par Next, we ascertain the validity of the proposed plan $\sigma^\prime$ by checking each district is contiguous. If $\sigma^\prime$ is not a valid districting, it is rejected and the process is repeated for $\sigma_t$. 
    \par Finally, this algorithm applies a Metropolis-Hastings filter to determine a probability of transitioning to $\sigma^\prime$ by setting $\sigma_{t+1} = \sigma^\prime$. The algorithm accepts the proposed plan as $\sigma_{t+1}$ with a probability proportional to its desirability. To do this, we use the weight function $w(\sigma)$, where the probability of accepting $\sigma^\prime$ as $\sigma_{t+1}$ is 
    \[
    Pr(\text{accepts $\sigma^\prime$ as $\sigma_{t+1}$}) = \begin{cases}
    1 &\text{ if } w(\sigma^\prime) >  w(\sigma_t)\\
    \frac{w(\sigma^\prime)}{w(\sigma_t)} & \text{ if }w(\sigma^\prime) \leq  w(\sigma_t)
    \end{cases}.
    \] As stated above, we define our weight function as
    \[
    w(\sigma) = e^{-(\beta_1pop\_eq(\sigma) + \beta_2 comp(\sigma))}.
    \]

    The Flip algorithm continues this process until completing its target number of accepted steps, where an accepted step is the case where the algorithm transitions to a proposed plan, setting $\sigma_t = \sigma^\prime$.  
    
\subsection{Reproducing a study in North Carolina}
    \par 
    In order to investigate the validity of the Redist flip algorithm, we set about reproducing the results of Herschlag et al. \cite{herschlag_quantifying_2020}. For this experiment, we use the same voting, population, and shapefile data, retrieving these data sets from the US Census Bureau \cite{bureau_tigerline_nodate}. We will be using the 2010 Voting Tabulation Shape File for North Carolina and the 2016 US House of Representatives Voting data. We then set about selecting parameters that would produce the most reliable and efficient simulations.
    \par First, we looked at restricting our plans to an acceptable population range for each district. In compliance with Supreme Court ruling \textit{Karcher v. Daggett} \cite{kd_karcher_1983}, the population tolerance of Congressional Districts should be equal unless the state can provide good reason for the unbalance. Despite this ruling, running the flip algorithm without a population tolerance is not feasible, as the acceptance rate becomes negligibly close to 0, making movement of the chain near impossible. For this reason, we select a population tolerance of  $10 \% $, as it provides the smallest population variance while still complying with \textit{Brown v. Thompson}, which allows for a $10\%$ variance in State and local districting elections \cite{altman_redistricting_nodate}. 
    \par 
    Next, we select a constraint weight for compactness. This parameter, however, was selected almost entirely by its affect on running time. Since compactness itself is not precisely defined in political science, we simply aimed to avoid plans that overtly violate compactness. We found that a compactness score of .4 results in a manageable running time, while providing plans that satisfy a loose visual compactness metric. The measuring of compactness itself is a difficult and problematic field, and is discussed in further detail by Roland Fryer and Richard Holden \cite{fryer_measuring_2007}.
    \par 
    Third, we select the number of simulation steps to run for each sampled plan. Since there is no theoretical justification of a mixing time for this chain, we must resign ourselves to using approximation metrics to determine the mixing of the chain. We used the Gelman-Rubin diagnostic of convergence to obtain a minimum number of simulation steps required. Using 10 independent chains, we increased the number of simulation steps until we received a Gelman-Rubin less than 1.1, as is the indication of a well-mixing chain  \cite{gelman_inference_1992}. With this metric, we decided to run 3000 simulation steps in our recreation of North Carolina for each sampled plan. We ran this simulation for 6 days to obtain 50 sampled plans, serving as a small ensemble of plans for North Carolina. 
\begin{table}[h!]
\centering
\begin{tabular}{ |c|c|c|c|c| } 
 \hline
 Pop Tolerance & Constraint  & Weight & Nsims & Nplans \\ 
 \hline
 0.1 & Edges Removed & 0.4 & 3000 & 50\\ 
 \hline
\end{tabular}
\caption{North Carolina Redist Flip Parameters}
\end{table}

Finally, we calculated the number of Democratic seats won in the 2016 House of Representatives Vote, plotting the distribution of Democratic seats won to compare to results of Herschlag et al. \cite{herschlag_quantifying_2020}.
    
\subsection{Generating a Simulated Annealing Algorithm}
Improving the speed of generating each sampled plan is satisfied through simulated annealing. Working with the 2020 Iowa County level shapefile, we designed an algorithm to run simulation steps without any constraint restriction before slowly annealing back to the population and compactness constraints described above. We choose to use the Iowa County shapefile as a basis for design because of its simplistic nature, allowing for fast experimentation on the cooling schedule, or rate at which we reintroduce the population and compactness constraints. 

\par First, we run 500 steps without constraint weights, which we refer to as hot steps. These steps contribute to the majority of the mixing of the chain, and since we are running without constraints, we can complete them quickly with a 100\% acceptance rate. When running 1000 hot steps, we see a similar distribution of sampled plans, allowing us to loosely conclude that 500 hot steps is sufficiently mixing the chain. 
\par 
To achieve the same desired distribution as the flip algorithm with constraints, we need to re-introduce the constraint parameters at a rate that prevents a near-zero acceptance rate. Reintroducing the constraint too quickly will prevent the chain from accepting any proposed plan, resulting in a completely halted, or stagnant chain.  

\par With this goal in mind and after experimentation, we decided to choose a population tolerance change of 0.005, or .5\%. That is, we would run a flip chain for 10 steps before decreasing the population tolerance by 0.05. Starting with a population tolerance of 1, we would repeat this procedure until arriving at our target population tolerance of .1, or 10\%.  With these parameter choices, we would require 1800 annealing steps for population alone. 
\par 
Similarly, starting with a compactness weight of 0, we increase our compactness constraint weight by 0.01 until reaching our desired weight of 0.4. The result is 400 steps to reintroduce the desired compactness weights. The choice to anneal population and compactness separately, along with the cooling schedule were chosen to minimize the number of steps necessary to return to desirable constraint weights, while avoiding with likely probability the case of a stagnant chain. 
\begin{table}[h!]
\centering
\begin{tabular}{ |c|c|c|c|c| } 
 \hline
 Hot Steps & Anneal Steps & Cold Steps & Pop Tolerance delta  & Compact delta\\ 
 \hline
 500 & 10 & 100 & 0.005 & 0.01\\ 
 \hline
\end{tabular}
\caption{Simulated Annealing Parameters}
\end{table}

\par 
Finally, we run the flip algorithm for 100 steps with our desired constraints, which we will call cold steps. The purpose of these 100 steps is not to provide any further mixing, but instead to ensure that our resulting plan has an acceptable compactness and population score. 
    
\par
After configuring our simulated annealing algorithm, we tested it in North Carolina. We compare a distribution composed of 300 sampled plans from our annealing algorithm to the distribution produced by Herschlag et al. \cite{herschlag_quantifying_2020}.

\subsection{Expanding to Texas} 
Proceeding in a similar fashion to that of North Carolina, we set about generating an ensemble for Texas Congressional districting plans. We retrieve the 2020 Texas VTD shapefile and population data from the Redistricting Data Hub, which was collected by the US Census Bureau. We also retrieve the 2020 US House of Representatives election also from the US Census Bureau \cite{bureau_tigerline_nodate}. 
\par We then set about preprocessing the 2020 US House data file to align with our Texas VTD shapefile. We iterate through each county of Texas, matching precinct names recorded in the voting data to VTD names recorded in our population data. In the case that multiple precincts fall into the same VTD, we sum the precinct votes and record it as the vote for the VTD. Once completing this matching procedure by name, we matched the remaining unmatched precincts based on geographic data through a Google maps query. After assigning all precinct voting data to a VTD, we still had 9 out of 9007 VTDs which had not yet been assigned a vote from the election. These VTD's were extremely sparsely populated regions, with all having a population less than 10. The majority (6 of 9) of these problematic VTD's were located in County 29 and 439, while the other 3 were located in County 37, 41, and 415.  We assigned both the Democratic and republican vote of these VTDs as half the population of the VTD.

Using the same constraints as in North Carolina, or population tolerance of 10 \% and compactness weight of 0.4, we choose the minimum number of simulation steps required to achieve a Gelman-Rubin score less than 1.1 when running 10 separate chains. We arrive at 5000 necessary steps, and sample 50 plans both with a compactness constraint and without a compactness constraint. 

\begin{table}[h!]
\centering
\begin{tabular}{ |c|c|c|c|c| } 
 \hline
 Pop Tolerance & Constraint  & Weight & Nsims & Nplans \\ 
 \hline
 0.1 & Edges Removed & 0.4 & 5000 & 50\\ 
 \hline
\end{tabular}
\caption{Texas Redist Flip Parameters}
\end{table}
\par
The parameters for the chains creating this ensemble can be found in \text{Table 5}. While more than 50 sampled plans is preferable, the amount of time required to produce each individual time dictated our 50 plan total.

\par 
Taking the same approach for our simulated annealing algorithm in Texas as in North Carolina, we generated 500 plans using the same parameters shown in \text{ 3}.
 
\par 
Finally, we plotted the number of Democratic seats won for each ensemble of plans representing our distributions, comparing to the number of seats won in the existing plan used in the 2020 US House of Representatives Vote. 

\section{Discussion}
 
\par We found that the 2020 Texas Congressional Districting gave an unfair advantage to Republicans in the 2020 US House election. The results shown in \textbf{Figure 5} demonstrate an abnormal outcome in this election, which not only favors Republicans unfairly but is also too statistically improbable to be unintentional. Our distribution indicates that the enacted plan election outcome is 3 standard deviations from our generated mean, suggesting that fair selection of a such a plan is infeasible.  These findings not only convey manipulation within the redistricting system that conventional metrics of gerrymandering detection have failed to effectively address, but exhibit the 2020 Texas US House congressional election outcome as an extreme outlier of feasible election outcomes.

\par We emphasize that we do not see this study as adequately concluding on the degree to which Texas has an unfair congressional districting. Our validation study has shown that insufficient samples may lead to distorted probabilities on election outcomes. Seeing how we were not able to generate a large enough sample size from which we created our reference distribution, we understand that our 0.0002 probability the 2020 US House election outcome with a new districting plan are not representative. The small sample size is likely to significantly decrease this approximated likelihood, producing an inaccurate quantification of gerrymandering in Texas. 
\par However, we did produce 550 near-random viable congressional districting plans for Texas without a single one yielding the result seen in the actual election. This is clear evidence indicating that a Congressional districting plan yielding 13 Democratic seats must be intentionally manipulated for the benefit of the Republican party. While we do not argue that a fair plan would yield our mean of 17 Democratic seats won, we believe such a drastic difference should inspire further investigation into the extent to which such a result is unfair. With study as motivation, we hope to see further investigation into larger elections with recognition that Simulated Annealing is a promising method allowing political analysis of larger geographic areas. 

\bibliographystyle{alpha}
\bibliography{mybib1.bib}

\end{document}